\documentclass[aps,prb,twocolumn,showpacs]{revtex4-1}
\usepackage[latin1]{inputenc}
\usepackage{soul}
\usepackage{amsmath}
\usepackage[T1]{fontenc}
\usepackage{graphicx}
\usepackage{epstopdf}
\usepackage[LGRgreek]{mathastext}
\usepackage{graphicx}
\usepackage{grffile}
\usepackage{textcomp}
\usepackage{gensymb}
\usepackage{longtable}
\usepackage{lipsum}
\usepackage{array}

\begin{document}
\title{Robust spin liquid state against magnetic-dilution in the bi-layer Kagome material Ca$_{10}$Cr$_7$O$_{28}$} 
\author{Ashiwini Balodhi, Anzar Ali, and Yogesh Singh}
\affiliation{Indian Institute of Science Education and Research (IISER) Mohali,
Knowledge City, Sector 81, Mohali 140306, India}

\begin{abstract}
Recently, the bi-layer Kagome lattice material Ca$_{10}$Cr$_7$O$_{28}$ has been shown to be a quasi-two-dimensional quantum spin liquid (QSL) where the frustration arises from a balance between competing ferromagnetic and antiferromagnetic exchange within a bi-layer.  In an attempt to understand what happens when this balance is disturbed, we present a magnetic dilution study.  Specifically, we have synthesized Ca$_{10}$(Cr$_{1-x}$V$_x$)$_7$O$_{28}$ (0 $\leq$ x $\leq$ 0.5) where magnetic Cr$^{5+}$ ($S = 1/2$) is partially replaced by non-magnetic V$^{5+}$ ($S = 0$).  We also synthesized the fully non-magnetic isostructural material Ca$_{10}$V$_7$O$_{27.5}$.  We report a detailed structural, magnetic and heat capacity study on these materials.  A monotonic increase in the unit cell parameters is found for the Ca$_{10}$(Cr$_{1-x}$V$_x$)$_7$O$_{28}$ materials with increasing $x$.  An order of magnitude decrease in the Curie-Weiss temperature from $4$ to $0.5$~ K is found for the partial V substituted samples, which indicates a relative increase in antiferromagnetic exchange with increase in V content.  However, despite this change in the relative balance in the exchange interactions and the large disorder introduced, no magnetic ordering or spin-glass state is observed down to $2$~K in the V substituted samples.  The QSL state of the parent compound thus seems surprisingly robust against these large perturbations.

 \end{abstract}
\maketitle
\section{Introduction}
Although the Quantum spin liquid (QSL) state was first conceptualized by Anderson in 1973 when he was worrying about the ground state of the triangular lattice Heisenberg antiferromagnet \cite{anderson1973}, the discovery and study of QSLs received a boost after Anderson's proposal that the parent compounds of the Cuprate high-$T_c$ superconductors might be a "Resonating Valance Bond (RVB)" QSL \cite{anderson1987}.  The study of Quantum Spin-Liquids (QSL's) and geometrically frustrated magnets in general have since been attracting continuous attention, not only for discovering new QSL realizations but also for the possibility of finding superconductivity and other novel metallic states by tuning the QSL by doping or pressure  \cite{leon}$^-$\cite{Broholm}. 

The preferred recipe to construct spin liquids has been to put small spins on lattices made up of triangular motifs, having antiferromagnetic interactions between the spins.  The geometric frustration then suppresses the tendency to order and a QSL may be engineered.  This strategy has led to the realization of several QSL candidates including the quasi-two-dimensional triangular lattice organics , the quasi-2D Kagome material Herbertsmithite, and the 3D hyperkagome magnets like SrCr$_8$Ga$_4$O$_{19}$ (SCGO), Na$_4$Ir$_3$O$_8$, and PbCuTe$_2$O$_6$ \cite{leon, Savary,Wen,Zhou2017,Knolle,Broholm}.  

Recently a new QSL candidate Ca$_{10}$Cr$_7$O$_{28}$ has been discovered, where the QSL state is stabilized by a novel mechanism of competing ferromagnetic and antiferromagnetic exchange, with the FM exchange dominating \cite{balz}.  The crystal structure of Ca$_{10}$Cr$_7$O$_{28}$ is built up of distorted Kagome bi-layers stacked along the $c$-axis.  Within the bi-layers, each layer has two kinds of distorted equivalent triangles of Cr$^{5+}$ ($S = 1/2$) exist with triangles having ferromagnetic exchange surrounded by triangles having anti-ferromagnetic exchange, and the other way around. The bi-layers are magnetically isolated from other bi-layers as evidenced by inelastic neutron scattering \cite{balz}.  The frustration arises from the weak ferromagnetic exchange between the two layers within a bi-layer.  A combination of bulk and microscopic probes of magnetism established that Ca$_{10}$Cr$_7$O$_{28}$ has no magnetic order or spin-freezing down to $19$~mK and that below $0.4$~K the spins fluctuate at a rate independent of temperature \cite{balz}.  A broad and diffuse continuum of excitations was revealed in inelastic neutron scattering, consistent with spinon excitations \cite{balz}.  The Hamiltonian for this system has been arrived at by a spin wave analysis of excitations out of a fully polarized state \cite{balz2017a} and a pseudo- fermion functional renormalization group calculation confirmed a spin liquid state for this Hamiltonian \cite{balz}.  Although, no excitation gap was observed within experimental resolution in the INS measurements \cite{balz}, a recent thermal conductivity study suggested a gapped QSL state for Ca$_{10}$Cr$_7$O$_{28}$ \cite{Ni2018}.

These experimental discoveries have motivated several theoretical studies including a tensor-network method revealing absence of magnetic order and importance of quantum correlations \cite{Kshetrimayum}, and semiclassical theories using Monte Carlo and molecular dynamics methods \cite{Biswas,Pohle}. 
 
We have been trying to understand how the QSL state evolves under the influence of perturbations which can change the relative importance of the competing ferro- and anti-ferromagnetic exchanges.  Our recent high pressure magnetic study showed that application of pressure of $1$~GPa enhanced the Weiss temperature by a factor of two, but did not lead to any magnetic ordering suggesting that the QSL state was quite robust \cite{ashiwini}.

In this work, we report the effects of magnetic dilution and negative chemical pressure on the QSL properties of Ca$_10$Cr$_7$O$_{28}$.   This is achieved by partially replacing magnetic Cr$^{5+}$ with non-magnetic V$^{5+}$ ions in Ca$_{10}$(Cr$_{1-x}$V$_x$)$_7$O$_{28}$ (0 $\leq$ x $\leq$ 0.5).  In addition to magnetically diluting the lattice and introducing random disorder, V substitution also acts as negative chemical pressure resulting in a unit cell volume expansion of $\approx 1.2~\%$ for complete replacement of Cr by V.  By studying the structure, magnetic, and thermal properties of these materials we aim to understand the effect each of these parameters (magnetic dilution, pressure, and disorder) have on the QSL properties of the parent material.

Additionally, we also report the realization of a perfect nonmagnetic analog Ca$_{10}$V$_7$O$_{27.5}$, which enables an accurate exclusion of the lattice heat capacities without any fitting. Such an advantage is absent in the previous reports on Ca$_{10}$Cr$_7$O$_{28}$ \cite{balz,ashiwini}. The systematic investigation of magnetic susceptibility and heat capacity analysis shows the absence of any long-ranged magnetic ordering or spin-freezing down to $2$~K indicating that the QSL state of the parent material is surprisingly robust against large perturbations.

\begin{figure}[ht]
	\includegraphics[width= 8 cm]{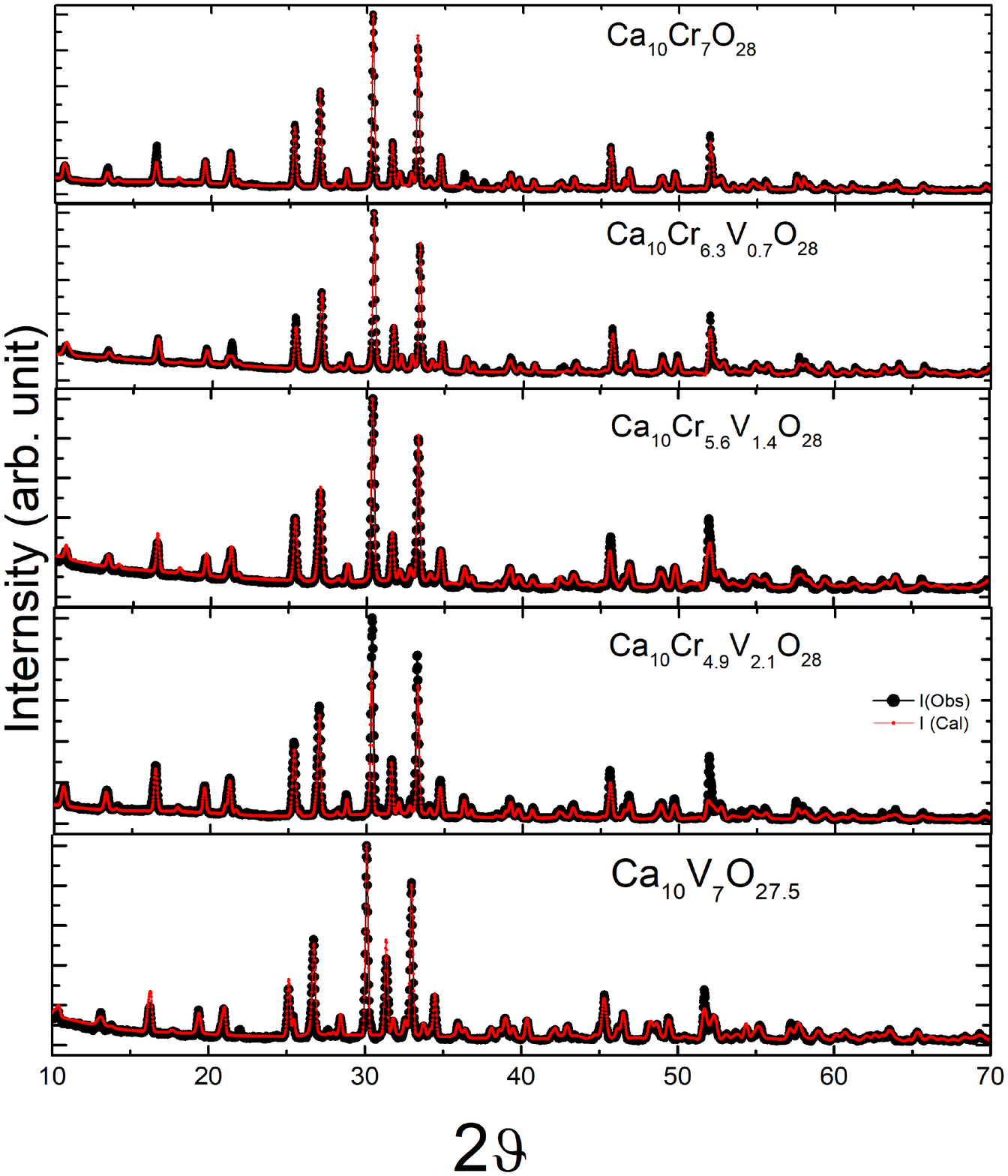}
	\caption{(Color online) Powder X-ray diffraction pattern for Ca$_{10}$(Cr$_{1-x}$V$_x$)$_7$O$_{28}$ (x = 0, 0.1, 0.2 and 0.3) and Ca$_{10}$V$_7$O$_{27.5}$. The solid circles represent the observed data, the solid lines through the data represent the fitted pattern.}
	\label{xrd}
\end{figure}

\begin{table*}[t]
	\centering
	\caption{ Structural parameters for Ca$_{10}$(Cr$_{1-x}$V$_x$)$_7$O$_{28}$  (x= 0, 0.1, 0.2, 0.3, 0.4 and 0.5, 1) obtained from a Rietveld refinement of room temperature powder x-ray patterns shown in Fig-\ref{xrd1} with space group 167, R3c.}
	
	\begin{tabular}  {c c c c c c c }	
		\hline \hline	
		Atom & Wyck & x & y& z& Occ. & B \AA  \\
		\hline
		
		Ca$_{10}$Cr$_{6.4}$V$_{0.7}$O$_{28}$, & a = b = 10.7978 \AA,& c=  38.175, \AA &  Cell volume = 4450.918 \AA$^3$    &   & & \\\\
		
		Cr1/V1  & 36f	& 0.3185	&0.1477       & 0.1324     &  0.95/ 0.035     & 0.0040    \\
		Cr2/V2  & 36f 	& 0.1744    &-0.1316      & -0.0992    & 0.92/0.068   	  & 0.00118  \\
		Cr3A  	& 12C   & 0.0000    & 0.0000      & -0.0994     & 0.73     	  	  & 0.025   \\
		Cr3B 	& 12C   & 0.0000    & 0.0000      & 0.5313     & 0.29		 	  & 0.025  \\\\
		 \hline
				
		Ca$_{10}$Cr$_{5.6}$V$_{1.4}$O$_{28}$, & a = b = 10.804 \AA,& c= 38.191 \AA, & Cell volume = 4457.816 \AA$^3$        & & & \\
		
		Cr1/V1  & 36f	& 0.3108	&0.1473       & 0.1324     &  0.84/ 0.15     & 0.009    \\
		Cr2/V2  & 36f 	& 0.1794    &-0.1420      & -0.0994    & 0.92/0.08   	  & 0.005  \\
		Cr3A  	& 12C   & 0.0000    & 0.0000      & -0.0064     & 0.72     	  	  & 0.002   \\
		Cr3B 	& 12C   & 0.0000    & 0.0000      & 0.0144     & 0.27		 	  & 0.025  \\\\
		
	\hline
		
		Ca$_{10}$Cr$_{4.9}$V$_{2.1}$O$_{28}$, & a = b = 10.811 \AA, &c =  38.201 \AA, &    Cell volume = 4464.7299 \AA$^3$       & &  & \\
		
		Cr1/V1  & 36f	& 0.3152	&0.1401      & 0.1342    &  0.81/0.189     & 0.0018    \\
		Cr2/V2  & 36f 	& 0.1738    &-0.1356      & -0.0988    & 0.9/0.1   	  & 0.011  \\
		Cr3A  	& 12C   & 0.0000    & 0.0000      & -0.0019     & 0.69     	  	  & 0.025   \\
		Cr3B 	& 12C   & 0.0000    & 0.0000      & 0.0195     & 0.32	 	  & 0.025  \\\\

		\hline
		
	Ca$_{10}$Cr$_{4.2}$V$_{2.8}$O$_{28}$, & a = b = 10.813 \AA &  c = 38.204 \AA, & Cell volume = 4466.637 \AA$^3$    &  && \\
		
		Cr1/V1  & 36f	& 0.3185	&0.1477       & 0.1324     &  0.95/ 0.035     & 0.0040    \\
		Cr2/V2  & 36f 	& 0.1744    &-0.1316      & -0.0992    & 0.92/0.068   	  & 0.00118  \\
		Cr3A  	& 12C   & 0.0000    & 0.0000      & 0.9938     & 0.77     	  	  & 0.025   \\
		Cr3B 	& 12C   & 0.0000    & 0.0000      & 0.5313     & 0.23		 	  & 0.025  \\\\
			
	\hline
		
	Ca$_{10}$Cr$_{3.5}$V$_{3.5}$O$_{28}$, & a = b = 10.820 \AA, &   c= 38.209(7) \AA, & Cell volume =  4473.384 \AA$^3$      & &  & \\
		
		Cr1/V1  & 36f	& 0.3185	&0.1437       & 0.1314     &  0.95/ 0.035     & 0.0040    \\
		Cr2/V2  & 36f 	& 0.1769    &-0.1376      & -0.0982    & 0.92/0.068   	  & 0.00118  \\
		Cr3A  	& 12C   & 0.0000    & 0.0000      & 0.9913     & 0.72    	  	  & 0.025   \\
		Cr3B 	& 12C   & 0.0000    & 0.0000      & 0.5173     & 0.28		 	  & 0.025  \\\\
	
	\hline
		
	Ca$_{10}$V$_7$O$_{28}$, & a = b = 10.853 \AA, &  c =38.228 \AA, &  Cell volume = 4502.939 (11) \AA$^3$       & & & \\
	
	V1  & 36f	& 0.31119	& 0.1405      & 0.1301     &  0.99    & 0.00709   \\
	V2  & 36f 	& 0.17807   &-0.1372      & -0.0999    & 1  	  & 0.0119  \\
	V3A & 12C   & 0.0000    & 0.0000      & 0.00515     & 0.73    & 0.0178   \\
	V3B & 12C   & 0.0000    & 0.0000      & 0.0403     & 0.29	  & 0.002987  \\
	O3A & 12C   & 0.0000    & 0.0000      & 0.9938     & 0.38    & 0.00178   \\
	O3B & 12C   & 0.0000    & 0.0000      & 0.5313     & 0.13	  & 0.002987  \\\\

	\hline
	\hline
	
	\end{tabular}
	\label{Table1}
	
\end{table*}

\section{Experimental details}
Polycrystalline samples of Ca$_{10}$(Cr$_{1-x}$V$_x$)$_7$O$_{28}$  ($x = 0, 0.1, 0.2, 0.3, 0.4, 0.5, 1$) were synthesized from CaCO$_3$, Cr$_2$O$_3$ and V$_2$O$_5$ powders ($99.99 \%$, Alfa-Aesar) by conventional solid state reaction method.  Stoichiometric amounts of starting materials were mixed thoroughly, loaded in alumina crucibles with a lid, and calcined at 900 $^o$C for $30$~h.  The resulting powders were then pressed into a pellet and heated at $1020~^o$C for $40$~hrs in air with an intermediate grinding and pelletizing step.  The final pellet of all samples were quenched in Argon. The Powder X-Ray diffraction (PXRD) was collected at room temperature using a Rikagu diffractometer (Cu K$\alpha$) and a Rietveld refinement of the PXRD patterns were performed using the GSAS software. Magnetization and heat capacity measurements were done using a Quantum Design physical property measurement system (QD-PPMS).

\section{Results} 

\begin{figure}[ht]
	\includegraphics[width=5.5 cm,angle=-90]{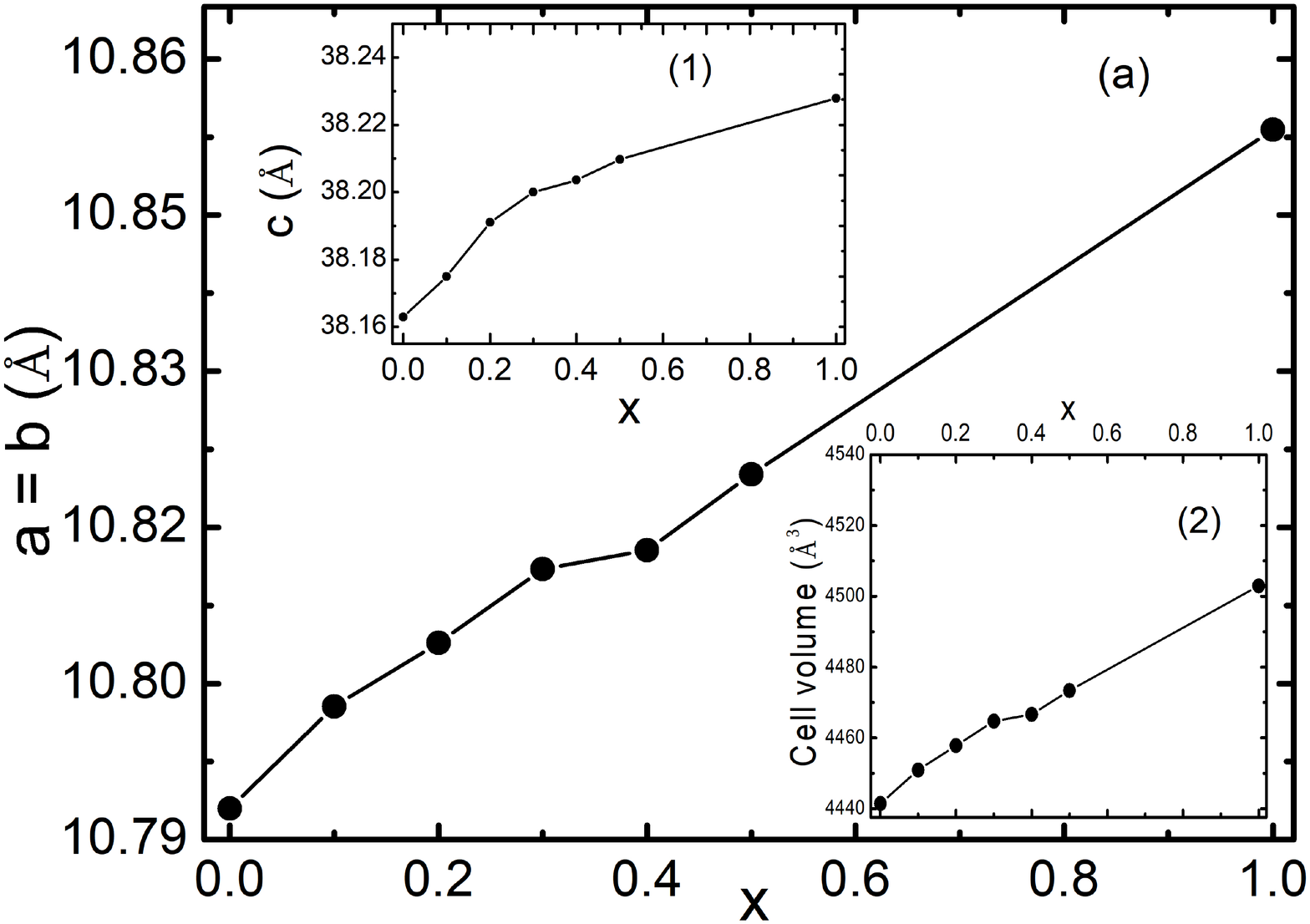}  
	\includegraphics[width=6.cm,angle=-90]{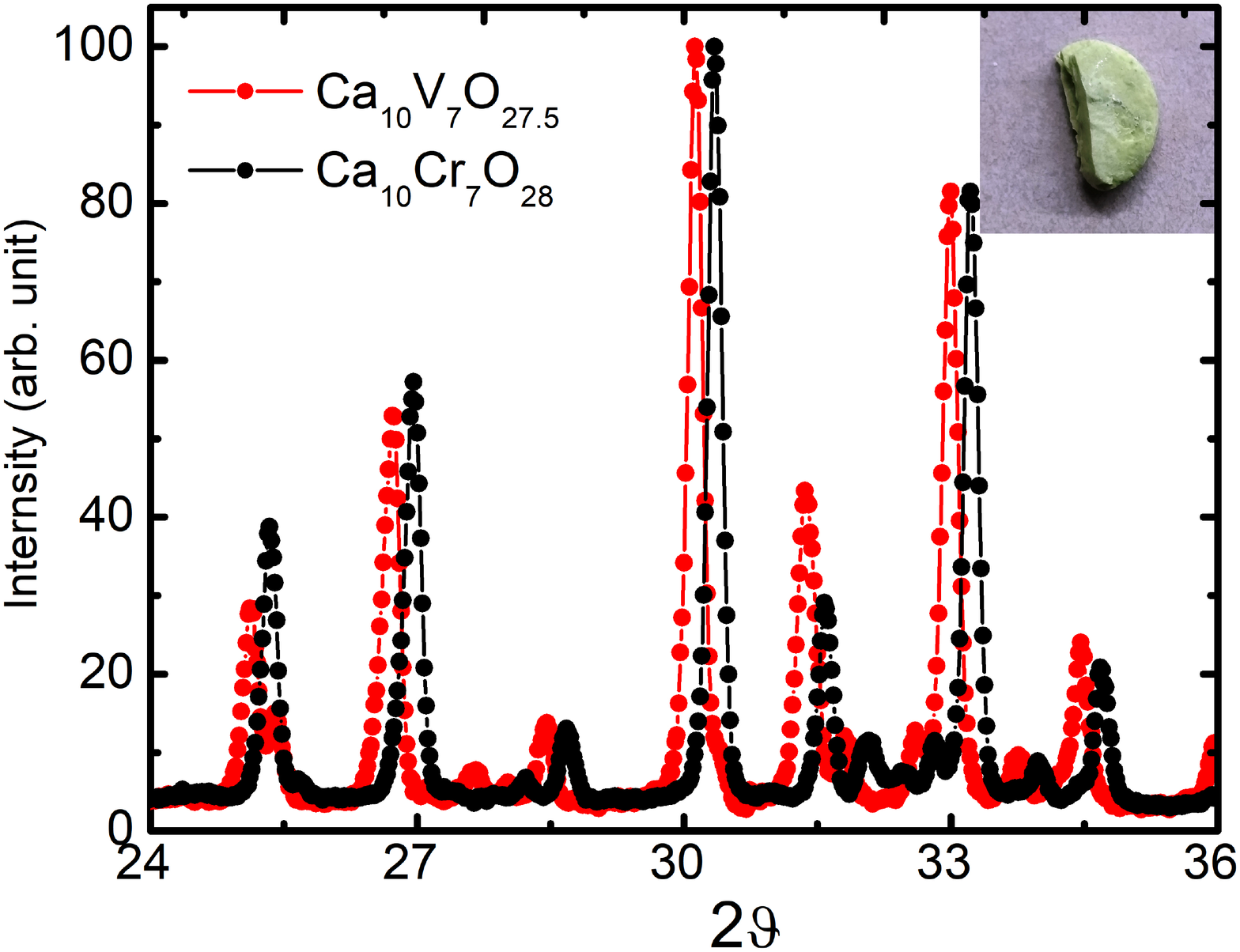} 
	\caption{(a) Variation of lattice constant "a=b" as a function of V  doping. Inset-I and Inset-II show variation of lattice constant "c"  and cell volume versus V  doping x respectively for Ca$_{10}$(Cr$_{1-x}$V$_x$)$_7$O$_{28}$ (0 $\leq$ x $\leq$ 0.5) and Ca$_{10}$V$_7$O$_{27.5}$ (b) Comparision of PXRD diffraction peak positions in intensity versus 2$\theta$ graph of Ca$_{10}$Cr$_7$O$_{28}$ (black) and Ca$_{10}$V$_7$O$_{27.5}$(red). A yellow green pellet of Ca$_{10}$V$_7$O$_{27.5}$ is shown on the up-right corner. }
	\label{xrd1}
\end{figure}

\subsection{ Powder diffraction Structure Analysis }
The crystal structure of Ca$_{10}$Cr$_7$O$_{28}$ is built up of distorted Kagome bilayers of Cr$^{5+}$ ($S = 1/2$ moments), stacked along the $c$-axis as described elsewhere \cite{balz,ashiwini}.  The room temperature PXRD data for Ca$_{10}$(Cr$_{1-x}$V$_x$)$_7$O$_{28}$ with different V concentrations $x = 0.0, 0.10, 0.2, 0.3$ along with a Rietveld refinement of the data are shown in Fig.~\ref{xrd}.  The PXRD data for all synthesized materials are found to be single-phase adopting the expected trigonal space group-R3c structure \cite{balz,ashiwini}.   The best fit parameters obtained from Rietveld refinements are listed in Table~\ref{Table1}.  There are multiple Cr sites in the parent structure and a substitution by V could lead to V being distributed randomly at any of the $3$ Cr sites.  However, we found a significantly better fit by making the natural assumption that the V$^{5+}$ ions would replace the Cr$^{5+}$ ions (Cr1 and Cr2 sites) instead of the Cr$^{6+}$ ions (Cr3A and Cr3B sites). The refined fractional occupation of Cr and V matched well with the target concentration of V as can be seen from the Table~\ref{Table1}.  Our analysis reveals an approximately linear increase in the lattice constant $"a"$ while the $"c"$ lattice parameter shows a weaker increase beyond about $x \geq 0.2$ leading to a monotonic increase in the cell volume as shown in the Fig.~\ref{xrd1}(a). This trend agrees well with expectation from a comparison of the radii of V$^{5+}$ (0.355 \AA) and Cr$^{5+}$ (0.345 \AA) ions.

The availability of an iso-structural non-magnetic material is always crucial to extract the magnetic contribution to the heat capacity of a magnetic material.  In previous reports, due to unavailability of such a material the magnetic heat capacity analysis of Ca$_{10}$Cr$_7$O$_{28}$ relied on approximate high-temperature fits \cite{balz,ashiwini}. Here we also report realization of a non-magnetic isostructural compound Ca$_{10}$V$_7$O$_{27.5}$, where we have succeeded in replacing Cr completely by V.  Since all vanadiums are in the V$^{5+}$ oxidation state, charge neutrality requires a slight oxygen deficiency from $28$ to $27.5$. PXRD data of Ca$_{10}$V$_7$O$_{27.5}$, is shown in the lowest panel of Fig.~\ref{xrd} and the absence of any extra diffraction peak, confirms the single-phase nature of the synthesized material. A comparison of the PXRD data for the $x = 0$ and $x = 1$ materials is shown in the lower panel of Fig.~\ref{xrd1} along with an image of a piece of the pellet for Ca$_{10}$V$_7$O$_{27.5}$.  As can be seen from Fig.~\ref{xrd1}, compared to the parent material, the PXRD for the $x = 1$ material is identical and shows a shifting of all Bragg peaks to lower angles suggesting an expansion of the lattice.  A refinement confirms this with an increase in cell volume of approximately $1.16~ \%$ as we increase the V  content from $x = 0$ to $x = 1$.  In addition to PXRD results, we successfully characterized the non-magnetic state of  Ca$_{10}$V$_7$O$_{27.5}$ by magnetic susceptibility and heat capacity measurements, which will be discussed in later sections.  All materials were found to be strongly insulating with two probe resistance values in the $M\Omega$ range at room temperature.

\section{Magnetic Susceptibility}

 \begin{figure}[ht]
	
\includegraphics[width= 8.25 cm]{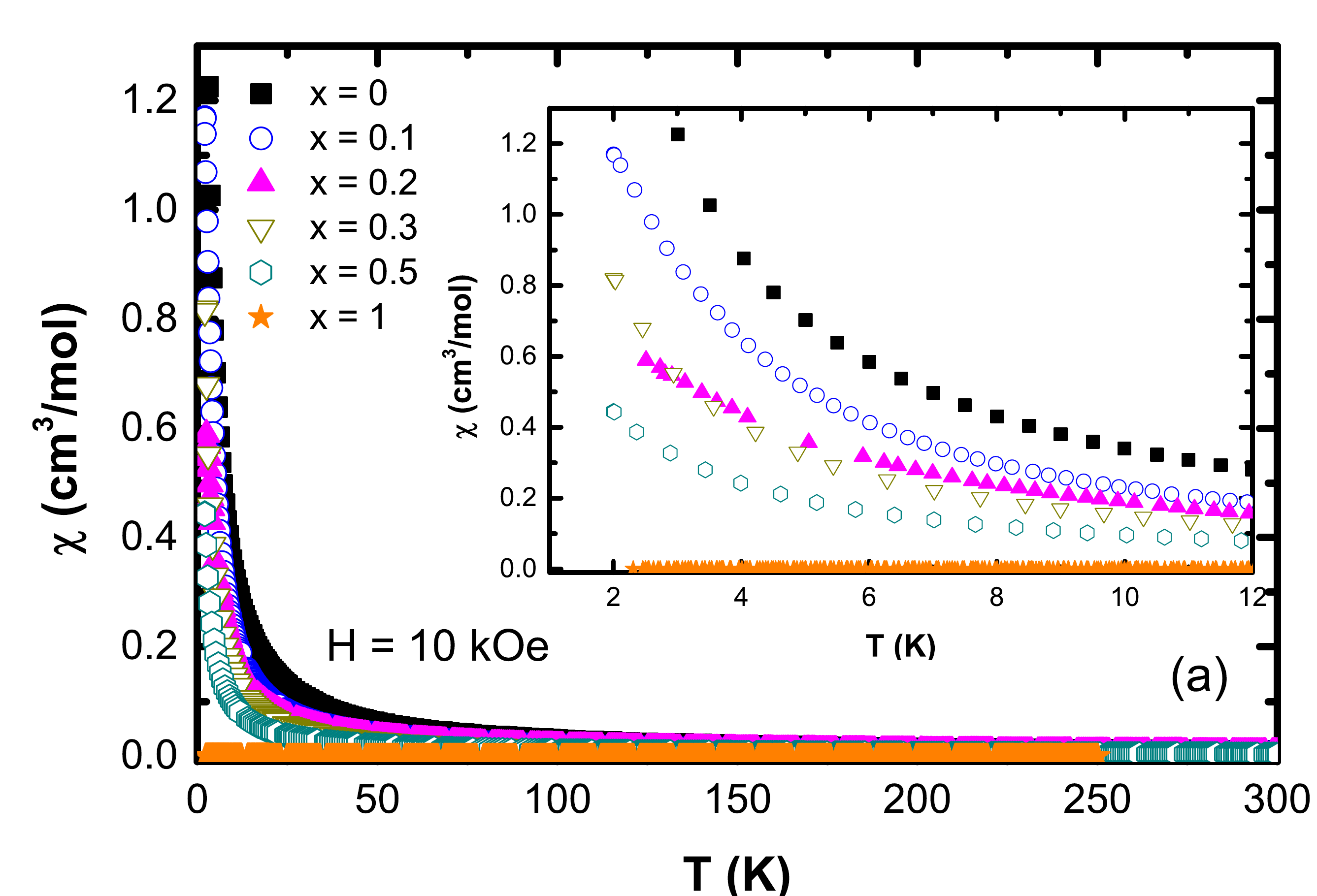}
\includegraphics[width= 8.25 cm]{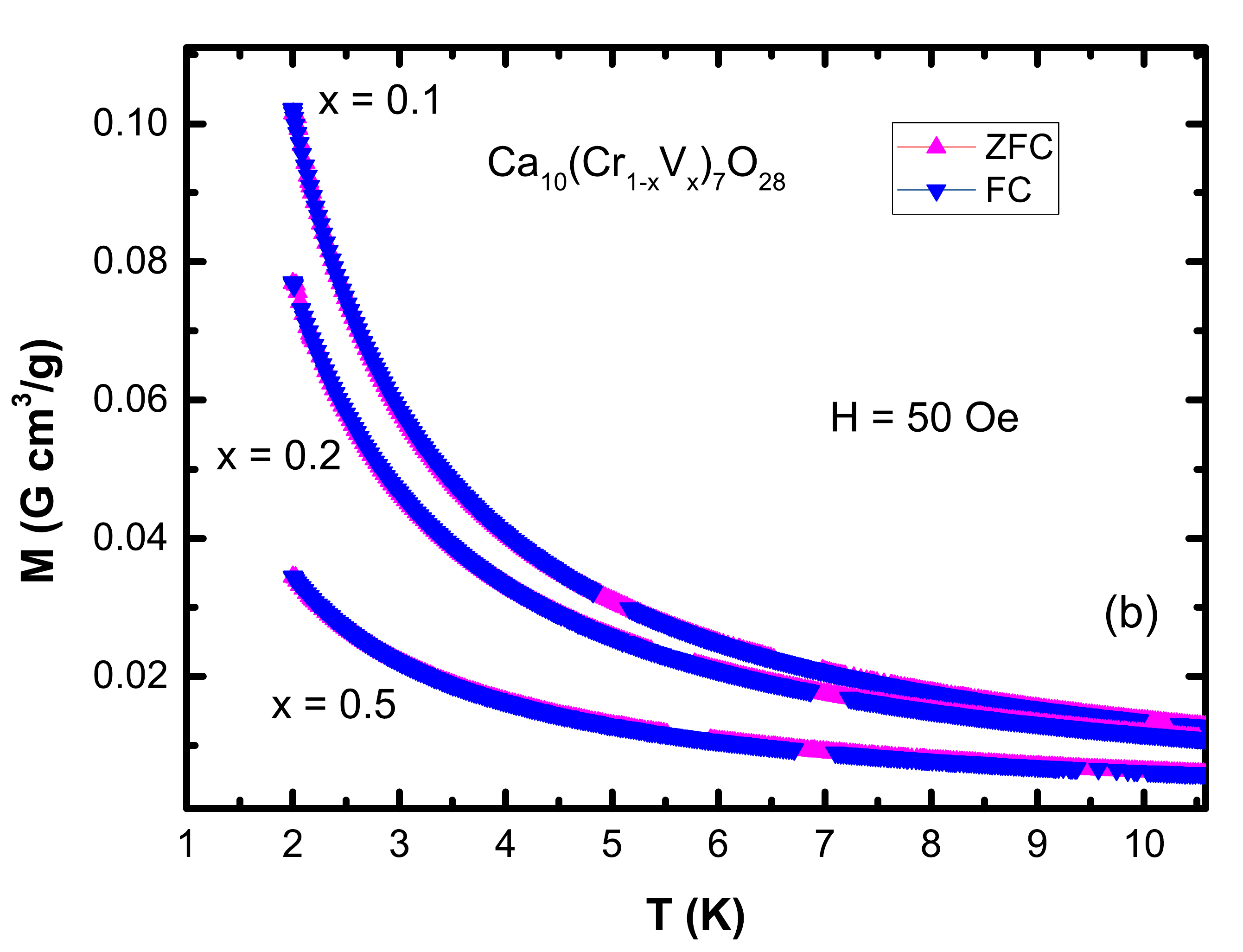}
\includegraphics[width= 8.25 cm]{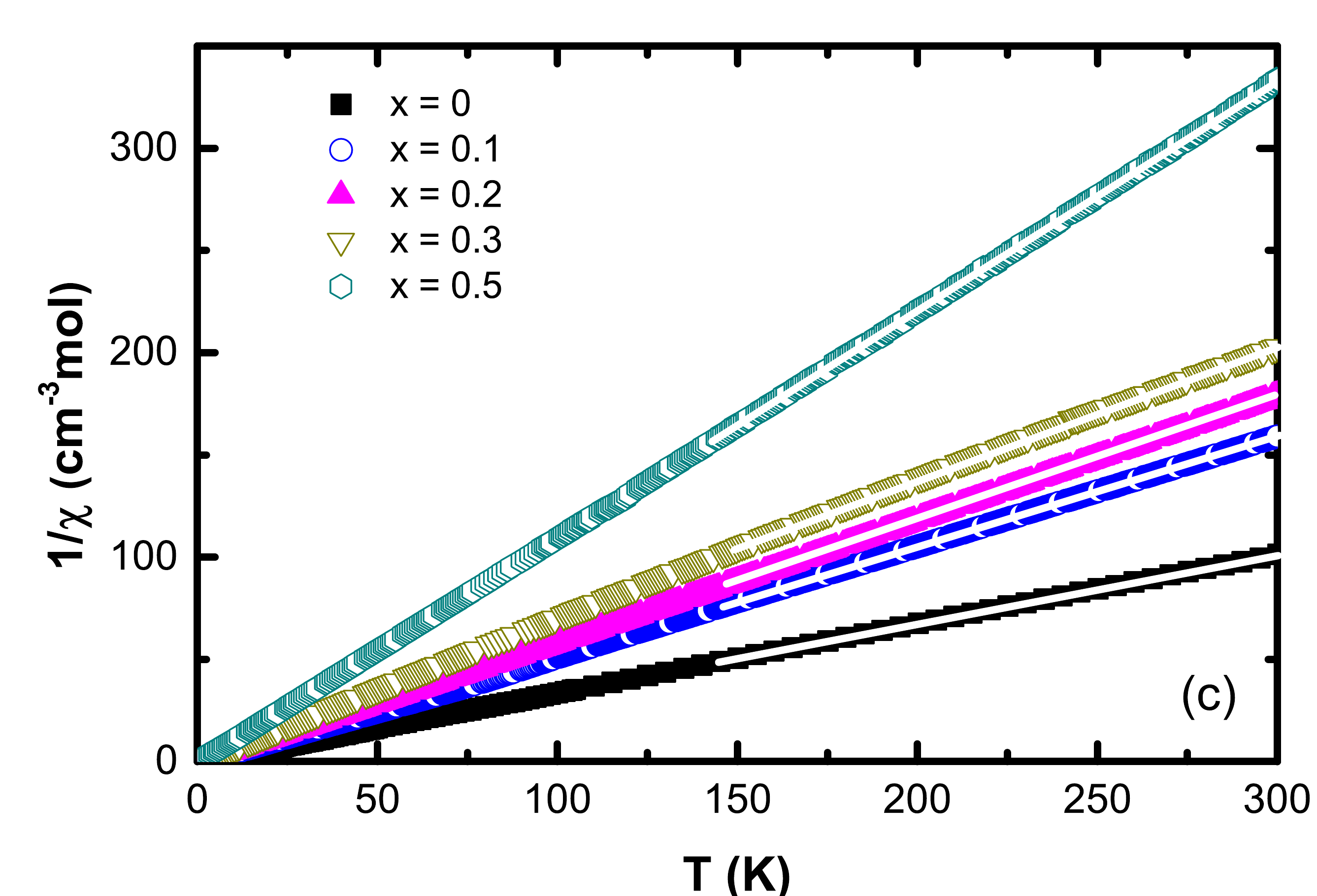}
\caption{ (color line) (a) Magnetic susceptibility $\chi$ versus temperature $T$ for Ca$_{10}$(Cr$_{1-x}$V$_x$)$_7$O$_{28}$ (0 $\leq$ x $\leq$ 0.5) and Ca$_{10}$V$_7$O$_{27.5}$ in a magnetic field $H = 1000$~Oe.  Inset shows the data below $12$~K\@.  (b) Low field ZFC and FC $\chi(T)$ data for $x = 0.1, 0.2, 0.5$ materials showing absence of any hysteresis.  (c) $1/\chi$ vs $T$ for Ca$_{10}$(Cr$_{1-x}$V$_x$)$_7$O$_{28}$ (0 $\leq$ x $\leq$ 0.5)} 
\label{vsm}
\end{figure}

DC magnetic susceptibility $\chi = M/H$ versus temperature at a magnetic field $H = 1000$~Oe for  Ca$_{10}$(Cr$_{1-x}$V$_x$)$_7$O$_{28}$ (0 $\leq$ x $\leq$ 0.5) and Ca$_{10}$V$_7$O$_{27.5}$ are shown in Fig.~\ref{vsm}(a).  The inset shows the data at low temperatures.  The $\chi$ decreases in general with increase in $x$ which supports the picture of non-magnetic V$^{5+}$ being substituted for magnetic Cr$^{5+}$.  The change in the absolute value of $\chi$ is however from a combination of a reduction of the number of magnetic ions in the material, as well as a change in the magnetic exchange interactions as we will show later, and therefore the reduction in $\chi$ is not monotonic with increase in $x$.   Inspite of the apparently large disorder introduced into the magnetic sub-system by the random substitution of non-magnetic V$^{5+}$ for magnetic Cr$^{5+}$, we do not find any signature of a spin-glass like state at low temperatures.  This can be seen in the absence of any history dependence in the low field ($50$~Oe)  zero-field-cooled and field-cooled magnetization measurements shown in Fig.~\ref{vsm}(b) for the $x = 0.1, 0.2, 0.5$ materials.

As discussed earlier, the PXRD data is consistent with V$^{5+}$ being substituted at the Cr$^{5+}$ site.  Therefore, to make a quantitative analysis of the magnetic data we re-write the chemical formula as Ca$_{10}$(Cr$^{5+}_{1-x}$V$^{5+}_x$)$_6$Cr$^{6+}$O$_{28}$.  The $1/\chi$ data for the $x = 0 - 0.5$ are shown in Fig.~\ref{vsm}(c) and are all approximately linear in $T$ at high temperatures.  The $1/\chi(T)$ for $T \geq 150$~K were fit by the Curie-Weiss expression $\chi = \chi_0 + C /T-\theta$ with $\chi_0$, $C$, and $\theta$ the fitting parameters and the fits are shown as the curves through the data in Fig.~\ref{vsm}(c).  The parameters obtained from these fits are given in Table~\ref{Table-IV}. The value of the Curie constant per magnetic Cr ion were estimated from the $1/\chi$ data assuming the chemical formula Ca$_{10}$(Cr$^{5+}_{1-x}$V$^{5+}_x$)$_6$Cr$^{6+}$O$_{28}$.  It can be seen from Table~\ref{Table-IV} that the value of $C$ for all the materials is within about $10\%$ of the value $0.375$ expected for spin $S = 1/2$ moments with a $g$-factor $g = 2$.   This again supports the fact that non-magnetic V$^{5+}$ ions replace the magnetic Cr$^{5+}$ ions in the proportion close to the target compositions.

Additionally, a monotonic reduction of the Curie-Weiss temperature $\theta$ is found which indicates that the ferromagnetic exchange weakens relative to the antiferromagnetic exchange with increase in V  content.  This trend is opposite to what was observed on the application of pressure, which we have previously shown leads to a significant increase in the Curie-Weiss temperature \cite{ashiwini}.   Structural analysis on the current samples has shown that V substitution leads to an increase in the cell parameters and cell volume, i.e. it acts as a negative pressure.  Thus the reduction in the Curie-Weiss temperature on V substitution can be viewed as mostly a pressure effect.

\begin{table}
	\begin{tabular}  {|c|c|c|c|}	
		\hline	
		& $\chi_0 \frac{ 10^{-5} cm^3}{(mol}$ & $\theta (K)$ & C $\frac{cm^3K}{Cr^{5+}}$ \\ \hline 
		
		x =   0 	& 4.31  	& 4.1(1)	& 0.417 \\ \hline 
		x = 0.1    	& 1.56 	& 2.6	(1)	& 0.344 \\ \hline 
		x = 0.2	& -1.4 	& 1.78(2)	& 0.355\\ \hline
		x = 0.3  	& 3.7 	& 1.3	(1)	& 0.327 \\ \hline 
		x = 0.5  	& -2.2	& 0.1(1)	& 0.323  \\ \hline 
	\end{tabular}
	\caption{Parameters obtained by Curie-Weiss fit of magnetic susceptibility of Ca$_{10}$(Cr$_{1-x}$V$_x$)$_7$O$_{28}$ (0 $\leq$ x $\leq$ 0.5) in the temperature range 100-400 K.}
	\label{Table-IV}
\end{table} 

In Fig.~\ref{vsm}(a), we have also shown the diamagnetic susceptibility of the end member material Ca$_{10}$$V_7$O$_{27.5}$ confirming that it is a completely non-magnetic material.  We will use this to extract the magnetic contribution to the heat capacity next.

\begin{figure}[t]
	\includegraphics[width=8.75 cm]{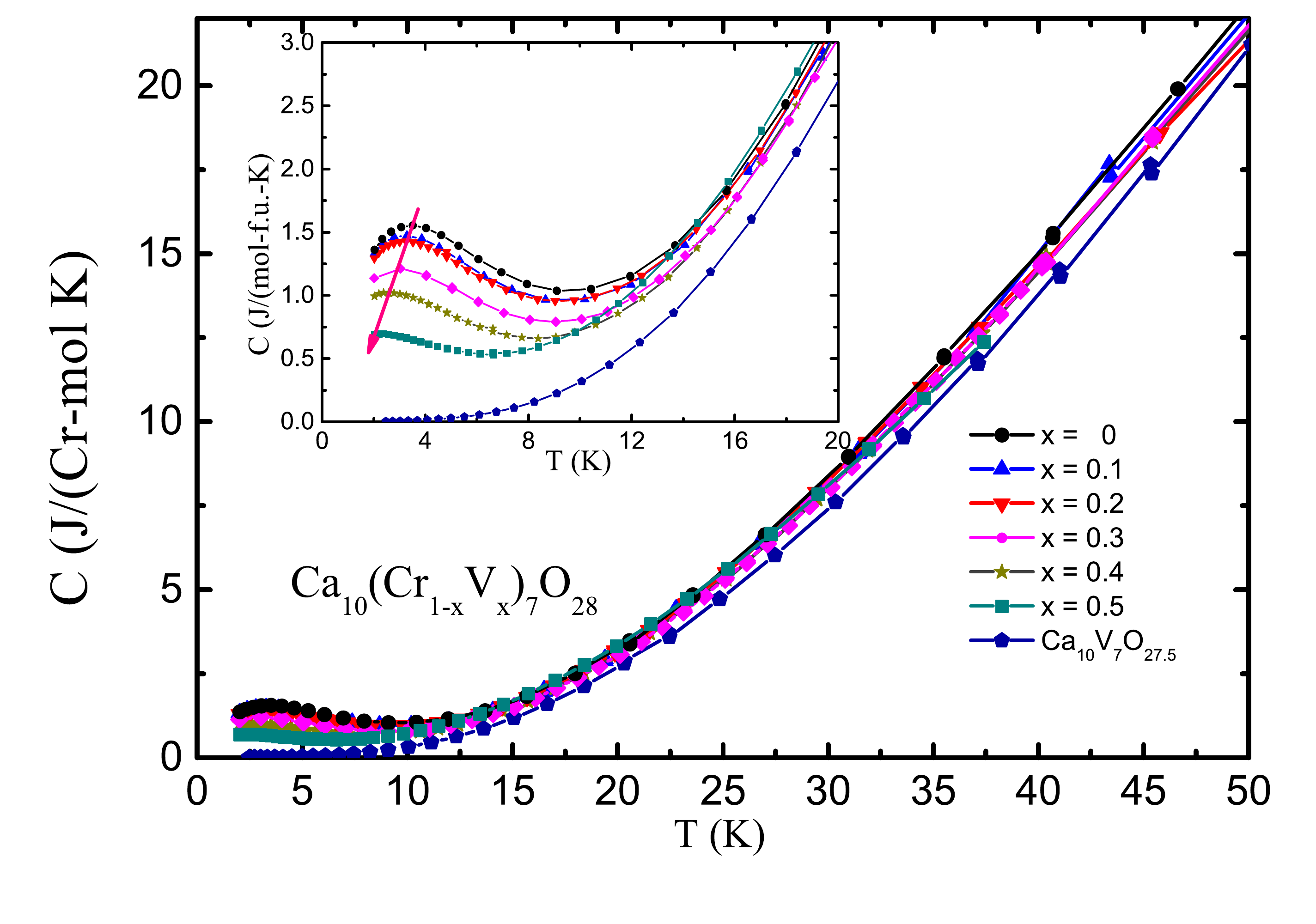}
	\caption{ (color line) (a) Heat capacity versus temperature T(K) for Ca$_{10}$(Cr$_{1-x}$V$_x$)$_7$O$_{28}$ (0 $\leq$ x $\leq$ 0.5) and Ca$_{10}$V$_7$O$_{27.5}$ up to 50 K. The Inset shows heat capacity versus temperature for all doped samples in the low-temperature range 2-20 K.} 
	
	\label{cp}
\end{figure}

The temperature dependence of the heat capacity of Ca$_{10}$(Cr$_{1-x}$V$_x$)$_7$O$_{28}$ (0 $\leq$ x $\leq$ 0.5) and Ca$_{10}$V$_7$O$_{27.5}$ are shown in Fig.~\ref{cp}.  The inset of Fig.~\ref{cp} shows the data at low temperatures. An upturn below $\sim 10$~K and a weak maximum around $T = 3$~K in Ca$_{10}$$Cr_7$O$_{28}$ is consistent with the previous reports \cite{balz, ashiwini}. This maximum was argued to signal a cross-over from a high temperature thermally fluctuating regime to a low temperature QSL regime with coherent quantum fluctuations \cite{balz} and the onset of short-ranged magnetic correlations.  From the inset we can see that the magnitude and peak temperature of this anomaly are approximately linearly suppressed with increasing V concentration.

\begin{figure}[t]
		\includegraphics[width=7.25 cm,angle=-90]{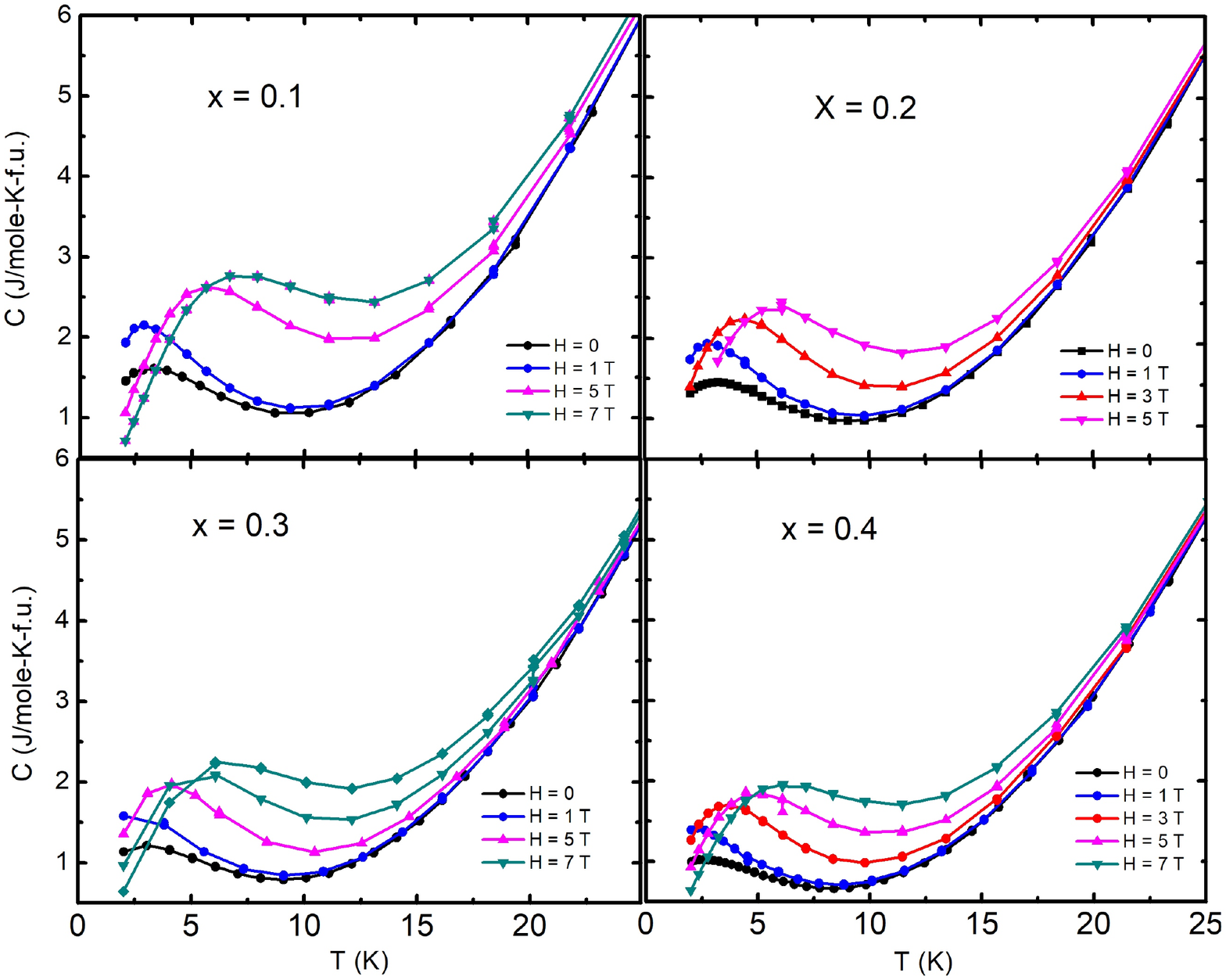} 
	\caption{(color line) Heat capacity versus temperature T(K) for Ca$_{10}$(Cr$_{1-x}$V$_x$)$_7$O$_{28}$ x = 0.1, 0.2, 0.3, and 0.4 at different applied magnetic fields H= 0-9 T.} 
	
	\label{cp1}
\end{figure}

For the parent compound Ca$_{10}$$Cr_7$O$_{28}$ it was shown that a magnetic field leads to a linear increase in the temperature of the maximum until a field of about $12$~T at which a fully polarized state is achieved \cite{ashiwini, balz2017a}.  
The magnetic field dependence of the low-temperature upturn for Ca$_{10}$(Cr$_{1-x}$V$_x$)$_7$O$_{28},~ x=0.1, 0.2, 0.3, 0.4$ are shown in Fig.~\ref{cp1}.  For all materials, the low temperature maximum moves to higher temperatures on the application of a magnetic field similar to the behaviour for the parent material.

\begin{figure}[h]
	\includegraphics[width=9 cm]{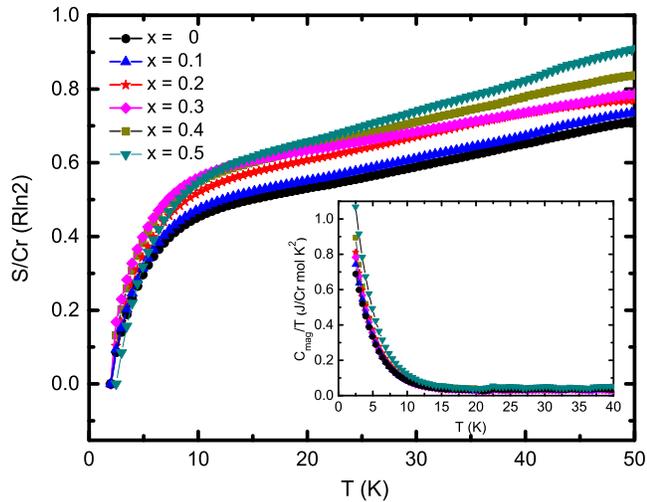}
	\caption{ (color line) Magnetic entropy per magnetic Cr $S$ (in units of Rln$2$) versus temperature $T$ for Ca$_{10}$(Cr$_{1-x}$V$_x$)$_7$O$_{28}$ (0 $\leq$ x $\leq$ 0.5).  Inset shows the magnetic heat capacity divided by temperature $C_{mag}/T$ versus temperature.} 
	\label{entropy}
\end{figure}

The availability of an iso-structural non-magnetic material in Ca$_{10}$V$_7$O$_{27.5}$ enables an accurate subtraction of the lattice heat capacity without the need for any approximate high temperature fitting which had to be used previously \cite{ashiwini}. The lattice contribution to the heat capacity for the magnetic materials was obtained by measuring the heat capacity of Ca$_{10}$V$_7$O$_{27.5}$ and then rescaling the data to account for the molecular mass difference between the magnetic samples and Ca$_{10}$V$_7$O$_{27.5}$.  The magnetic entropy $S$ per magnetic ion was then estimated by integrating the magnetic heat capacity $C_{mag}/T$ versus $T$ data.  The C$_{mag}/T$ versus $T$ is shown in Fig.~\ref{entropy} inset and the entropy $S$ estimated from these data for all the magnetic materials is shown in the main panel of Fig.~\ref{entropy}.   The entropy for all materials is qualitatively similar to the parent material with a rapid increase upto $\sim 10$~K followed by a weaker increase to the highest temperatures of measurement.  We note however, that none of the materials achieve the full entropy upto $50$~K, which is larger than the largest magnetic energy scales in this material.  This clearly suggests that similar to the parent material, small but significant amounts of entropy must be recovered for the $V$ substituted materials at still lower temperatures than the lowest temperature ($T = 2$~K) of our measurements.

\section{discussion and summary}
We have made a structural and magnetic property study of the effect of magnetic dilution on the QSL properties of the bi-layer Kagome material Ca$_{10}$Cr$_7$O$_{28}$.  This is in continuation of our previous high pressure work on Ca$_{10}$Cr$_7$O$_{28}$ \cite{ashiwini} and is an attempt to understand how the QSL state evolves on tuning away from the materials natural structure and composition.

The Vanadium substitution for Cr in Ca$_{10}$Cr$_7$O$_{28}$ has multiple effects; a pressure effect, a magnetic dilution effect, and a disorder effect.  From our structural refinement of the powder x-ray diffraction patterns of Ca$_{10}$(Cr$_{1-x}$V$_x$)$_7$O$_{28}$, we found a monotonic increase of the unit cell parameters.  This means that V substitution for Cr leads to an expansion of the lattice, or a negative pressure effect.   
Our previous study on the evolution of magnetic susceptibility with pressure had shown that the Curie-Weiss temperature increased from $4$~K to $7$~K under a pressure of $1$~GPa~ \cite{ashiwini}.  The Vanadium substitution studied here leads to a negative pressure and  therefore should be expected to reverse the trend seen in the high pressure measurements.  This is indeed observed as we find a decrease of the Curie-Weiss temperature from $4$~K to $\approx 0$~K for the Ca$_{10}$(Cr$_{1-x}$V$_x$)$_7$O$_{28}$ materials in going from $x = 0$ to $x = 0.5$.  Our studies therefore show that with the magnetic energy scales being small in Ca$_{10}$Cr$_7$O$_{28}$, pressure and chemical pressure are quite effective in tuning the relative importance of ferro-magnetic and antiferromagnetic exchange interactions.  This can be seen from a large $\sim 10$~K change in the Weiss temperature that can be achieved between pressure and chemical negative pressure effects in a system with a largest magnetic energy scale of about $20$~K~ \cite{balz,ashiwini}.  Inspite of this large change in the Curie-Weiss temperature, we observe no signatures of any magnetic ordering or freezing in the studied materials.  This strongly suggests that the QSL state is robust against significant changes in the relative magnetic exchange interactions.  

Vanadium substitution also introduces random disorder into the magnetic sublattice of Ca$_{10}$Cr$_7$O$_{28}$.  Large disorder is expected to lead to a frozen or glassy state.  However, somewhat surprisingly our low field ZFC and FC magnetic susceptibility measurements did not show any signatures of magnetic irreversibility, indicating the absence of a frozen or glassy state at least down to $2$~K even for the largest disorder samples $x = 0.4, 0.5$. 

Finally, the magnetic dilution seems to weaken the frustration effect seen for the parent material.  This can be seen by the suppression of the broad magnetic anomaly in the magnetic heat capacity.  Also, from the magnetic entropy shown in Fig.~\ref{entropy} it is clear that more and more of the expected Rln$2$ entropy is recovered between $2$~K and $50$~K as $x$ increases.  Since magnetic frustration tends to push the entropy to be released at lower temperatures, the recovery of magnetic entropy at higher temperatures demonstrates a weakening of frustration for the Vanadium substituted materials.

\end{document}